# Experimental Evidence for Crossed Andreev Reflection

D. Beckmann* and H. v. Löhneysen[¶§]

*Forschungszentrum Karlsruhe, Institut für Nanotechnologie, P.O. Box 3640, D-76021 Karlsruhe, Germany
[§]Forschungszentrum Karlsruhe, Institut für Festkörperphysik, P.O. Box 3640, D-76021 Karlsruhe, Germany
[¶]Physikalisches Institut, Universität Karlsruhe, D-76128 Karlsruhe, Germany

**Abstract.** We report on electronic transport properties of mesoscopic superconductor-ferromagnet spin-valve structures. Two ferromagnetic iron leads form planar tunnel contacts to a superconducting aluminum wire, where the distance of the two contacts is of the order of the coherence length of the aluminum. We observe a negative non-local resistance which can be explained by crossed Andreev reflection, a process where an electron incident from one of the leads gets reflected as a hole into the other, thereby creating a pair of spatially separated, entangled particles.

**Keywords:** superconductivity, ferromagnetism, Andreev reflection, entanglement.
**PACS: 74.45.+c, 03.67.Mn, 85.75.-d**

## INTRODUCTION

We have recently reported on the experimental investigation of electronic transport properties of superconductor-ferromagnet non-local spin-valve structures [1]. On the length scale of the superconductor's coherence length, spin-dependent transport was observed at subgap bias voltages. Our data were explained with a model based on the superimposition of two processes, namely crossed Andreev reflection (CAR) and elastic cotunneling (EC). However, our experimental setup and resolution were not sufficient to delineate the contribution of these two processes. Here, we report on preliminary data of our next generation experiment which overcomes these limitations, and show evidence for dominating CAR at low bias voltages.

## EXPERIMENT

Our sample layout consists of two ferromagnetic iron leads A and B which form tunnel contacts to a weakly oxidized superconducting aluminum wire (Fig. 1a). The contact separation is a few 100 nm, comparable to the coherence length of the aluminum wire. Contact A is used to inject a DC current $I_A$, while contact B measures the voltage $U_B$ with respect to the chemical potential of the superconductor. A spin-up electron incident from contact A on the superconductor in the source-drain voltage window from 0 to $U_A$ can be transmitted to contact B either as spin-up electron at positive energy (EC) or as spin-down hole at negative energy (CAR) (see Fig 1b).

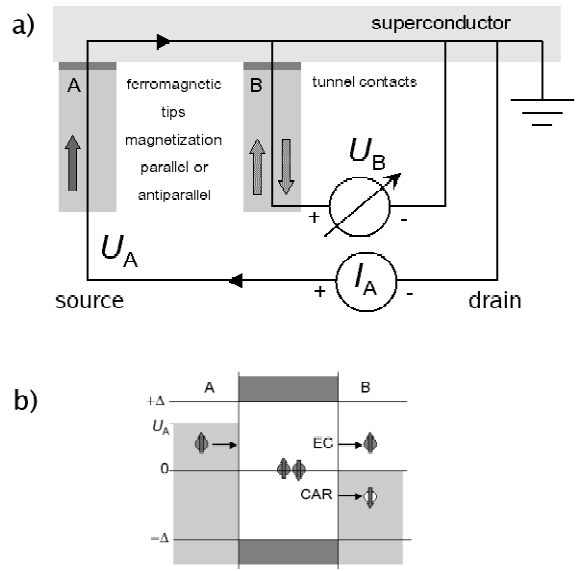

**FIGURE 1.** Experimental scheme: a) Ferromagnetic leads A and B form tunnel contacts to a superconducting bar. A is used for current injection, B for voltage detection inside the current path. b) Energy scheme for CAR and EC (see text).

For EC (or incoherent electron transmission in the normal state, including the effects of spin accumulation), the voltage $U_B$ is therefore always inside the source-drain window, i.e. for positive $U_A$ also $U_B$ will be positive. For CAR, $U_B$ will then be negative, i.e. outside the source-drain window [2]. This issue has been discussed in Ref. 2 in a different setup with only local Andreev reflection, but applies to our situation as well.

## RESULTS

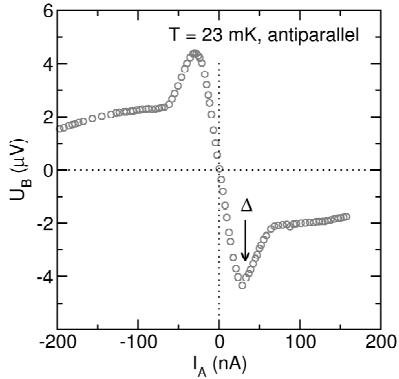

**FIGURE 2.** Bias dependence of $U_B$. Below the energy gap of the superconductor, a negative voltage (i.e. outside the source-drain window) is observed.

Figure 2 shows the non-local voltage $U_B$ as a function of injector current $I_A$ for one of our samples at low temperature for antiparallel magnetization alignment, where CAR is favored over EC due to the reversed spin of the hole. At low positive bias currents, a negative voltage is observed, i.e. $U_B$ is outside the source-drain window. At higher bias current, the slope of $U_B$ becomes positive. The turnaround occurs at the current which corresponds to $U_A = 200\mu V$, i.e. at the superconducting energy gap of aluminum, as indicated by the arrow in Fig. 2. Similar behavior was seen for several samples. For one sample, we observed a dominating positive slope at low bias, followed by a negative slope at higher bias (but still below the gap), similar to the observations made by Russo et al. [3] in a different experimental setup using an AC method, as opposed to our DC experiment. The reason for the qualitatively different behavior of some samples is subject to ongoing investigations.

## DISCUSSION

Our previous experiment [1] featured a non-local voltage detection (i.e. outside the current path), which has the advantage of being extremely sensitive to both spin accumulation and coherent non-local processes. However, in such a setup the sign of the non-local voltage is not conclusive evidence for CAR, as one compares two different voltages inside the source-drain window. In the presence of spin accumulation, the measured voltage can have either sign even in the normal state without CAR. The most significant change over our previous experiment is therefore the detection of the voltage $U_B$ inside the current path, where the observation of a negative $U_B$ for positive $U_A$ conclusively means that $U_B$ is outside the source-drain window, indicating CAR as the dominating non-local process. The positive slope at bias voltages above the superconducting energy gap can be attributed to the onset of electron transmission through allowed quasiparticle states.

## CONCLUSION

We have shown the observation of an unusual negative four-probe resistance occurring in superconductor-ferromagnet spin-valve-like structures. The effect can be explained by crossed Andreev reflection (CAR). Our results show that it may be feasible to create solid-state entanglers with CAR as the dominating transport process. Further systematic investigation is required for a better understanding of CAR compared to competing processes like elastic cotunneling.

## ACKNOWLEDGMENTS

We thank D. Feinberg and R. Melin for useful discussions, and especially P. Samuelsson, D. Sanchez, R. Lopez, E. Sukhorukov and M. Büttiker for bringing the source-drain window argument to our attention. This work was partly supported by the Deutsche Forschungsgemeinschaft within the Center for Functional Nanostructures.